\documentclass[journal]{new-aiaa}
\usepackage[utf8]{inputenc}
\usepackage{textcomp}
\usepackage{comment}
\usepackage{graphics}
\usepackage{graphicx}
\usepackage{hhline}
\usepackage{multirow}
 
 \usepackage{amsthm}
\usepackage{amsmath}
\usepackage[version=4]{mhchem}
\usepackage{siunitx}
\usepackage{multicol}
\usepackage{subfig}
\usepackage{longtable,tabularx}
\newtheorem{remark}{Remark}
\newtheorem{theorem}{Theorem}
\newtheorem{lemma}{Lemma}
\newtheorem{property}{Property}
\newtheorem{proposition}{Proposition}
\newtheorem{definition}{Definition}
\title{Cooperative Target Capture using Voronoi Region Shaping}

\author{Gautam Kumar \footnote{Research Scholar, Department of Aerospace Engineering; gautamkumar1@iisc.ac.in.} and Ashwini Ratnoo \footnote{Associate Professor, Department of Aerospace Engineering; ratnoo@iisc.ac.in. Associate Fellow AIAA }}
\affil{Indian Institute of Science, Bengaluru 560 012, India}

\begin{document}

\maketitle


\section{Introduction}

Multiplayer pursuit-evasion (MPE) scenarios have gained a lot of attention for their application in the area of missile-target interception \cite{dong2021three}, border-defence problem \cite{von2020multiple}, and territory protection from aerial intruders \cite{chakravarthy2020collision}. One of its variants involves multiple pursuers and a single evader/target. Therein, the pursuers use strategies to capture an evader while the evader's job is to evade successfully or to maximize the capture time.

The earliest of the works considers a single pursuer-single evader differential game in an unbounded domain where both players employ optimal strategies to achieve their objectives \cite{isaacs1967differential}. A border defence differential game is considered in \cite{garcia2019optimal,yan2017defense} where the pursuer's task is to capture the evader infiltrating the region of interest. This problem is later explored for the MPE game in \cite{garcia2020multiple} where optimal strategies are devised for the pursuers to capture the evaders before they enter the restricted region. The MPE problem of capturing a single evader is formulated as a target-attacker-defender scenario in \cite{wei2018optimal} where a group of pursuers lures the target to the reachable set of the remaining pursuers such that an encirclement around the target is achieved. The pursuit strategies in \cite{garcia2020multiple,isaacs1967differential,garcia2019optimal,yan2017defense,wei2018optimal} require the solution of the Hamilton-Jacobi-Isaacs (HJI) equation. With multiple players in the game, the non-uniqueness of the terminal condition complicates the synthesis of strategies for the players in that approach \cite{yan2019task}. Solving the HJI equation in an MPE game scenario is computationally expensive and realizing any cooperation among agents adds further complexity.

To overcome these limitations, various geometric approaches are presented in the literature. Ref. \cite{kopparty2005framework} presented a framework for capturing a target using multiple pursuers while assuming that the target is initially inside the convex hull of the pursuers. A trapping chain formation-based strategy is proposed in \cite{bopardikar2008discrete} where a bounded rectangular region is swept by a chain of UAVs leading to encirclement and capture of target. The capture of a faster evader is considered in \cite{chen2016multi} by employing a fishing strategy in which the pursuers surround the target and then shrink the encirclement region leading to the capture of the evader. That strategy requires a minimum number of pursuers and a specific initial position distribution of the pursuers around the target. Apollonius circle-based pursuit strategies for capturing a constant speed evader are investigated in \cite{ramana2017pursuit,sun2022cooperative} wherein a minimum number of pursuers are required to encircle the evader at the initial time. A relay pursuit strategy for capturing a single target using multiple pursuers is proposed in \cite{pan2022region} where a moving pursuer uses a policy that reduces the distance between the target and the stationary pursuers. The geometric approaches discussed in \cite{bopardikar2008discrete,chen2016multi,ramana2017pursuit,sun2022cooperative,pan2022region} consider the speed of the evader and pursuers to be constant.

Due to its simple construction, efficient partitioning of the region of interest, and properties that enhance cooperation among agents, the idea of Voronoi partitioning is employed in several aerospace applications such as path planning \cite{upadhyay2021voronoi}, airspace sectorization \cite{xue2009airspace}, and area coverage \cite{lum2010search,dong2016cooperative}. A relay pursuit strategy is proposed in \cite{bakolas2012relay} wherein the region is dynamically divided using Voronoi partitioning and the pursuers are assigned targets based on the minimum capture time. Voronoi partition-based techniques are also used in MPE scenarios. A class of geometric approaches \cite{huang2011guaranteed,zhou2016cooperative,pierson2016intercepting,shah2019grape,wang2023distributed} utilize Voronoi partitioning to define the evader's proximity region as the Voronoi cell whose generator is the evader's position. An area minimization policy for multiple pursuers-single evader problem is presented in \cite{huang2011guaranteed} where a guaranteed finite time capture of the evader is established, and that work is further explored for a non-convex environment in \cite{zhou2016cooperative}. In \cite{pierson2016intercepting,shah2019grape}, a monotonic decrease in the area of the evader's proximity region is achieved by a move-to-centroid strategy which directs a pursuer towards the center of the boundary of the Voronoi edge shared between that pursuer and the evader. For capturing an evader in an unbounded domain, the evader is first brought within the convex hull of the pursuers, and then the move-to-centroid strategy is used to deduce pursuers' control inputs for capturing the evader \cite{wang2023distributed}. In \cite{huang2011guaranteed,zhou2016cooperative,pierson2016intercepting,shah2019grape,wang2023distributed}, the pursuers' motion strategy is designed independently of the target's motion. Further, in those works, it is assumed that the evader moves with a speed less than or equal to that of the pursuers. Consequently, the shrinkage of the evader's proximity region is achieved by relying on the assumptions made rather than directly imposing the conditions on the system dynamics. This reflects a restrictive and indirect treatment of the problem. Additionally, the time of capture of the evader is not analyzed in the MPE strategies \cite{huang2011guaranteed,zhou2016cooperative,pierson2016intercepting,shah2019grape,wang2023distributed}.

In contrast, this paper directly considers the motion of the target and vertices of its proximity region, and proposes a motion policy that guarantees a monotonic decrease in the area of the evader's proximity region. The contributions of the work are as follows:
\begin{enumerate}
    \item Given an initial position distribution of pursuers, the Voronoi Diagram is employed to characterize the evader’s proximity region. The key idea is to dynamically shape that region using a policy that directs its vertices towards its instantaneous centroid. Analysis of the resulting dynamics of the evader's proximity region deduces the velocity control inputs for the pursuers.
    \item  Using the proposed motion policy, the evader’s proximity region is shown to shrink exponentially irrespective of its speed and evasion policy. As a result, the evader's capture is guaranteed without prior knowledge of the bounds on the evader's speed. Further, a conservative upper bound on the time of evader capture is deduced using the Chebyshev radius of the evader's proximity region.
    \item Validation studies show the capture of the evader while it moves with constant or variable speed. In one of the simulation cases, a second-order polynomial Kalman filter is employed to estimate the velocity and position of the evader using its noisy position information.
\end{enumerate}

The remainder of the paper is organized as follows: Section \ref{sec:2} introduces the problem. The evader region and its dynamics are analyzed in Section \ref{sec:3}. Section \ref{sec:4} presents the proposed motion policy which is demonstrated using simulations in Section \ref{sec:5}. Conclusions are drawn in Section \ref{sec:6}.

\section{Problem Statement}\label{sec:2}

Consider a two-dimensional region containing $(N+1)$ agents with $N (> 2)$ pursuers and $1$ evader. The position of $i$th pursuer and the evader is denoted by $\mathbf{x}^P_i(t) = [{x}^P_i,{y}^P_i]^T$ and $\mathbf{x}^E(t) = [{x}^E,{y}^E]^T$, respectively. All agents follow single integrator motion kinematics, that is,

\begin{equation}\label{eq:kinematics}
\begin{aligned}
    \Dot{\mathbf{x}}^P_i(t) &= u^P_i(t), ~ i \in \mathcal{N} =\{1,2,\dotsc,N\}\\
    \Dot{\mathbf{x}}^E(t) &= u^E(t),
\end{aligned}
\end{equation}
where $u^P_i, u^E \in \mathbb{R}^2$ denote $i$th pursuer's and the evader's velocity input, respectively. Further analysis considers the following assumptions:
\begin{enumerate}[label=(A\arabic*)]
    \item The initial position of the evader lies inside the convex hull formed by the set of pursuers' initial position. \label{assum:1}
    \item No two agents are initially collocated.
    \item The environment is obstacle-free.
    \item A centralized server knows agents' position and velocity information and relays the computed input to the pursuers.
\end{enumerate}

The pursuers' goal is to capture the evader while it follows an evasion policy unknown to the pursuer. The evader is captured at time $t_C \geq 0$ if the following condition satisfies:
\begin{equation}\label{eq:capture}
           \exists \; i:  ||\mathbf{x}^E(t_C)-\mathbf{x}^P_i(t_C)||  \leq r_c, i \in \mathcal{N},
\end{equation}
where $r_c\in \mathbb{R}^+$ is the radius of the capture zone of pursuers. The objective here is to deduce $u_i^P,~ i \in \mathcal{N}$ which leads to the capture of the evader.

\section{Evader Proximity Region}\label{sec:3}

The proximity region for an agent is defined as the set of all points in $\mathbb{R}^2$ which are closer to itself than to any other agent. Accordingly, the proximity region can be determined by \textit{Generalised Voronoi Diagram} (GVD). The GVD partitions $\mathbb{R}^2$ into sub-regions, $S(\mathbf{X}) $ where $\mathbf{X}$ is the set of generators which in this case is the set of agents' position, that is, $\mathbf{X} =\{\mathbf{x}^E,\mathbf{x}^P_1,\mathbf{x}^P_2,\dotsc, \mathbf{x}^P_N\}$.

\subsection{Agents' Proximity Region}

Given the set $\mathbf{X}$, the proximity region of $i$th agent, $S(\mathbf{X}_i)$ is defined as

\begin{equation}\label{eq:voronoi}
\begin{aligned}
    S(\mathbf{X}_i) =& \left \{\mathbf{x} : ||\mathbf{x} - \mathbf{X}_i|| \leq || \mathbf{x} - \mathbf{X}_j || \;   \right \},~~
    i,j=\{1,2,\dotsc, N+1\}, ~ i \neq j,~ \{\mathbf{X}_i,\mathbf{X}_j \} \in \mathbf{X}.
\end{aligned}
\end{equation}

\begin{property} \label{prop1} 
 \cite{preparata2012computational} $ S(\mathbf{X}_i) $ is a bounded convex region if $\mathbf{X}_i$ lies inside the convex hull formed by the elements in $\mathbf{X}$.
\end{property}
\begin{property} \label{prop2} 
 \cite{preparata2012computational} Any vertex point of $S(\mathbf{X}_i)$ is equidistant from two or more generators other than itself.
\end{property}

\begin{figure}[!hbt]
    \centering
    \includegraphics[width = 0.47\textwidth]{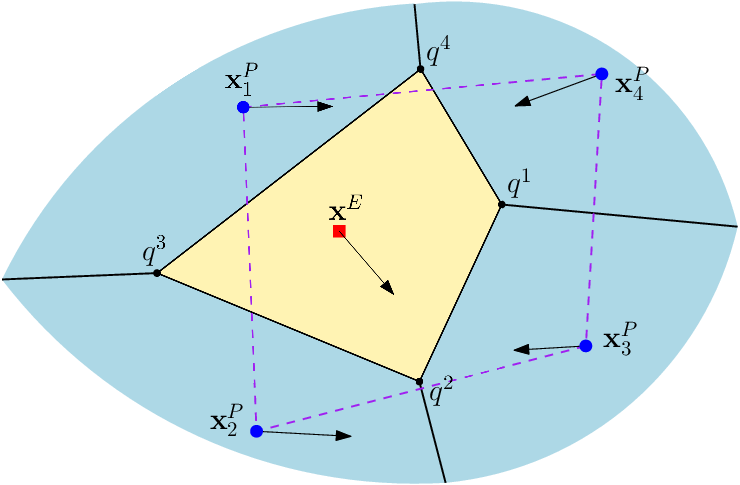}
    \caption{Voronoi partition with multiple agents in the region. (Square and circular markers indicate the position of the evader and pursuers, respectively.)}
    \label{fig:prob_form}
\end{figure}

An example scenario with 4 pursuers and 1 evader, and their proximity regions generated using (\ref{eq:voronoi}) is shown in Fig. \ref{fig:prob_form}. The dashed purple polygon is the convex hull formed by the agents' initial position. The yellow and blue regions represent the proximity region of the evader and the pursuers, respectively. The black solid lines represent the boundary of $S(\mathbf{X}_i),~\forall \mathbf{X}_i \in \mathbf{X}$. From \ref{assum:1} and Property (\ref{prop1}), $S(\mathbf{x}^E)$ is a bounded convex polygon.

\subsection{Evader Proximity Region Dynamics}

Consider the proximity region of the evader, $S(\mathbf{x}^E)$ in Fig. \ref{fig:prob_form}. The set of vertices of $S(\mathbf{x}^E)$ is denoted as $\mathcal{Q} = \{\mathbf{q}^1,\mathbf{q}^2,\dotsc, \mathbf{q}^l\}$, where $\mathbf{q}^i =[q^i_x,q^i_y]^T $. The vertices are arranged in a clockwise manner about the evader's position. Using the shoelace method for calculation of the area of a polygon from its vertices, the evader area $A_e$ is obtained as:
\begin{align}\label{eq:area_evader}
        A_e &=  \frac{1}{2}    \sum_{i =1}^l \left| \begin{array}{cc}
q^i_x & q^i_y \\
q^{i+1}_x & q^{i+1}_y 
\end{array}\right|,  ~(\text{where }\mathbf{q}^{l+1} = \mathbf{q}^{1})\\
\label{eq:area_evader1}
    &= \frac{1}{2}  [ 
(q^1_x q^2_y + q^2_x q^3_y + \dotsc + q^l_x q^1_y  )  - (q^1_y q^2_x + q^2_y q^3_x + \dotsc + q^l_y q^1_x  ) ] 
\\ 
&= \frac{1}{2}  \sum_{i =1}^l \left ( q^i_x q^{i+1}_y - q^{i+1}_x q^i_y \right ) \label{eq:area_evader2}.
\end{align}
Differentiating Eq. (\ref{eq:area_evader2}), the evader area dynamics is obtained as follows:

\begin{equation} \label{eq:area_evader_dot}
\begin{aligned}
     \Dot{A}_e &= \frac{1}{2}\sum_{i =1}^l \left ( \frac{\partial A_e }{\partial q^i_x } \Dot{q}^i_x + \frac{\partial A_e }{\partial q^i_y } \Dot{q}^i_y\right ), 
\text{ where }     \frac{\partial A_e }{\partial q^i_x } =\left (q^{i+1}_y - q^{i-1}_y \right ) ,\; \;  \; \;  \frac{\partial A_e }{\partial q^i_y } = \left(q^{i-1}_x - q^{i+1}_x\right ),\mathbf{q}^{l+1} = \mathbf{q}^1,\mathbf{q}^0 = \mathbf{q}^l.
\end{aligned}
\end{equation}
 \begin{remark}
     From Eq. (\ref{eq:area_evader_dot}), the rate of change of $A_e$ depends on the position and velocity of vertices of $S(x^E)$.
 \end{remark}

\subsection{Relation between Motion of Agents and Vertices of $S(\mathbf{x}^E)$}

The proximity region of all pursuers in $\mathcal{N}$ may not share boundary with $S(\mathbf{x}^E)$. By assumption (A1), the maximum number of such pursuers can be $N-3$. Consider $\mathcal{M}$ to be the set of $m (\leq N)$ pursuers whose proximity region shares a common boundary with $S(\mathbf{x}^E)$. Accordingly, the evader area dynamics in Eq. (\ref{eq:area_evader_dot}) is affected only by pursuers in $\mathcal{M}$ while the remaining pursuers are redundant and remain stationary. The pursuers in $\mathcal{M}$ are hereinafter referred to as active pursuers. 

\begin{remark}
    The number of vertices of $S(\mathbf{x}^E)$ equals the number of active pursuers, that is, $l=m$.
\end{remark}

\begin{figure}[!hbt]
    \centering
    \includegraphics[width = 0.5\textwidth]{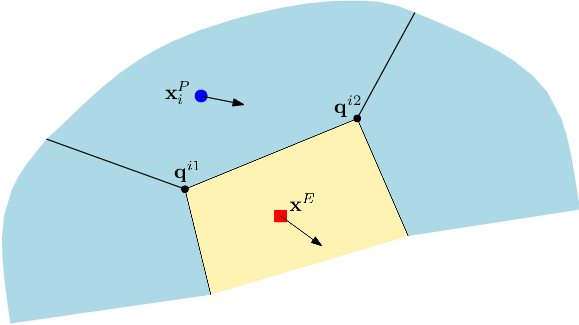}
    \caption{An example scenario depicting boundary shared between $S(\mathbf{x}^P_i)$ and $S(\mathbf{x}^E)$.}
    \label{fig:bpk}
\end{figure}

In Fig. \ref{fig:bpk}, $\mathbf{q}^{i1}=[q^{i1}_x,q^{i1}_y]^T$ and $\mathbf{q}^{i2}=[q^{i2}_x,q^{i2}_y]^T$ denote the end points of the edge shared between $S(\mathbf{x}^P_i)$ $(1\leq i \leq m)$ and the evader. The relation between $\{\mathbf{q}^{i1},\mathbf{q}^{i2}\}$ and $\mathbf{q}^j \in \mathcal{Q} (1 \leq j \leq l)$ is expressed as

\begin{align}\label{eq:q11_q1}
    \begin{aligned}
        \begin{bmatrix}
       \mathbf{q}^{11} \\
\mathbf{q}^{12}\\
\mathbf{q}^{21} \\
\mathbf{q}^{22} \\
\vdots\\
\mathbf{q}^{m1} \\
\mathbf{q}^{m2} 
    \end{bmatrix} = \begin{bmatrix}
       \mathbf{R}^{11} \\
\mathbf{R}^{12}\\
\mathbf{R}^{21} \\
\mathbf{R}^{22} \\
\vdots\\
\mathbf{R}^{m1} \\
\mathbf{R}^{m2}
    \end{bmatrix}\begin{bmatrix}
       \mathbf{q}^{1} \\
\mathbf{q}^{2}\\
\vdots\\
\mathbf{q}^{l} 
    \end{bmatrix},
    \end{aligned}
\end{align}
where $\mathbf{R}^{ik} =[\mathbf{r}^{ik}_1,\mathbf{r}^{ik}_2,\dotsc,\mathbf{r}^{ik}_l]$ $(k=\{1,2\})$ is defined as 
\begin{equation}
    \mathbf{r}^{ik}_j = \begin{cases}
        \begin{bmatrix}
            1 & 0\\
            0 & 1
        \end{bmatrix}, &\text{if } \mathbf{q}^{ik} = \mathbf{q}^j\\
        \mathbf{0}_{2\times 2}, &\text{otherwise.}
    \end{cases}
\end{equation}

Consider the example shown in Fig. \ref{fig:bpk}, Pursuer $P_i$ and Evader $E$ share the edge $\overline{\mathbf{q}^{i1} \mathbf{q}^{i2}}$. Using the Property \ref{prop2} of $S(\mathbf{x}^E)$, 
\begin{align}
  \label{eq:bpk1}  ||\mathbf{x}^P_i - \mathbf{q}^{i1}|| &= ||\mathbf{x}^E - \mathbf{q}^{i1}||, \text{ and} \\
    \label{eq:bpk2}||\mathbf{x}^P_i - \mathbf{q}^{i2}|| &= ||\mathbf{x}^E - \mathbf{q}^{i2}||. 
\end{align}

Following relations are obtained by squaring and then differentiating both sides of Eqs. (\ref{eq:bpk1}) and 
 (\ref{eq:bpk2}),
    \begin{align}
        (&\Dot{{x}}^P_i - \Dot{q}^{i1}_x) ({x}^P_i - q^{i1}_x) + (\Dot{{y}}^P_i - \Dot{q}^{i1}_y) ({y}^P_i - q^{i1}_y)=  (\Dot{{x}}^E - \Dot{q}^{i1}_x) ({x}^E - q^{i1}_x) + (\Dot{{y}}^E - \Dot{q}^{i1}_y) ({y}^E - q^{i1}_y) .\label{eq:bpk3} \\
        (&\Dot{{x}}^P_i - \Dot{q}^{i2}_x) ({x}^P_i - q^{i2}_x) + (\Dot{{y}}^P_i - \Dot{q}^{i2}_y) ({y}^P_i - q^{i2}_y)=  (\Dot{{x}}^E - \Dot{q}^{i2}_x) ({x}^E - q^{i2}_x) + (\Dot{{y}}^E - \Dot{q}^{i2}_y) ({y}^E - q^{i2}_y). \label{eq:bpk4}
    \end{align}

Rearranging terms in Eqs. (\ref{eq:bpk3}) and (\ref{eq:bpk4}), the velocity input to $P_i$, $u^P_i = [\Dot{{x}}^P_i, \Dot{{y}}^P_i]^T$ can be deduced in terms of position and velocity of the evader and vertices of $S(\mathbf{x}^E)$ as follows:

\begin{equation}\label{eq:bpk5}
    \begin{aligned}
         u^P_i 
 = A_i^{-1}B_i \begin{bmatrix}
\Dot{\mathbf{q}}^{i1} \\
\Dot{\mathbf{q}}^{i2}
\end{bmatrix}  + A_i^{-1}C_i \Dot{\mathbf{x}}^E,
    \end{aligned}
\end{equation}
where matrices $A_i \in \mathbb{R}^{2 \times 2}$, $B_i\in \mathbb{R}^{2 \times 4 }$, and $C_i \in \mathbb{R}^{2 \times 2 }$ are as follows:

\begin{align}\label{eq:coeff}
    \begin{aligned}
        A_i &= \begin{bmatrix}
        {x}^P_i - q^{i1}_x  & {y}^P_i - q^{i1}_y \\
         {x}^P_i - q^{i2}_x & {y}^P_i - q^{i2}_y
    \end{bmatrix},~
    B_i = \begin{bmatrix}
        {x}^P_i  - {x}^E  & {y}^P_i - {y}^E & 0 & 0 \\
         0 & 0 & {x}^P_i  - {x}^E & {y}^P_i - {y}^E
    \end{bmatrix},~
    C_i   =\begin{bmatrix}
         {x}^E -q^{i1}_x &{y}^E - q^{i1}_y \\
         {x}^E - q^{i2}_x & {y}^E - q^{i2}_y
    \end{bmatrix}.
    \end{aligned}
\end{align}
Differentiating Eq. (\ref{eq:q11_q1}) and using that in Eq. (\ref{eq:bpk5}) leads to

\begin{align}\label{eq:rel}
    u^P_i 
 = A_i^{-1}B_i \begin{bmatrix}
{\mathbf{R}}^{i1} \\
{\mathbf{R}}^{i2}
\end{bmatrix} \begin{bmatrix}
\Dot{\mathbf{q}}^{1} \\
\Dot{\mathbf{q}}^{2}\\
\vdots \\
\Dot{\mathbf{q}}^{l}
\end{bmatrix}  + A_i^{-1}C_i \Dot{\mathbf{x}}^E.
\end{align}
Using Eq. (\ref{eq:rel}), the velocity input for active pursuers can be expressed as
\begin{align}\label{eq:allpursuer}
    \begin{bmatrix}
        u^P_1\\
u^P_2\\
\vdots\\
u^P_m 
    \end{bmatrix} &=  \begin{bmatrix}
        A_1^{-1} B_1 & \mathbf{0}_{2\times 4}  &\dotsc & \mathbf{0}_{2\times 4}\\
        \mathbf{0}_{2\times 4} & A_2^{-1}B_2  & \dotsc &\mathbf{0}_{2\times 4}\\
        \vdots & \vdots & \ddots & \vdots\\
        \mathbf{0}_{2\times 4} & \mathbf{0}_{2\times 4} & \dotsc &A_m^{-1}B_m
    \end{bmatrix}  \begin{bmatrix}
       \mathbf{R}^{11} \\
\mathbf{R}^{12}\\
\mathbf{R}^{21} \\
\mathbf{R}^{22} \\
\vdots\\
\mathbf{R}^{m1} \\
\mathbf{R}^{m2}
    \end{bmatrix}\begin{bmatrix}
       \Dot{\mathbf{q}}^{1} \\
\Dot{\mathbf{q}}^{2}\\
\vdots\\
\Dot{\mathbf{q}}^{l} 
    \end{bmatrix}  + \begin{bmatrix}
        A_1^{-1} C_1 & \mathbf{0}_{2\times2}  &\dotsc & \mathbf{0}_{2\times2}\\
        \mathbf{0}_{2\times2} & A_2^{-1}C_2  & \dotsc &\mathbf{0}_{2\times2}\\
        \vdots & \vdots & \ddots & \vdots\\
        \mathbf{0}_{2\times2} & \mathbf{0}_{2\times2} & \dotsc &A_m^{-1}C_m
    \end{bmatrix}\begin{bmatrix}
        \Dot{\mathbf{x}}^E \\
\Dot{\mathbf{x}}^E \\
\vdots\\
\Dot{\mathbf{x}}^E  
    \end{bmatrix}.
    \end{align}

\section{Proposed Motion Policy}\label{sec:4}

\begin{figure}[!hbt]
    \centering
    \includegraphics[width=0.35\textwidth]{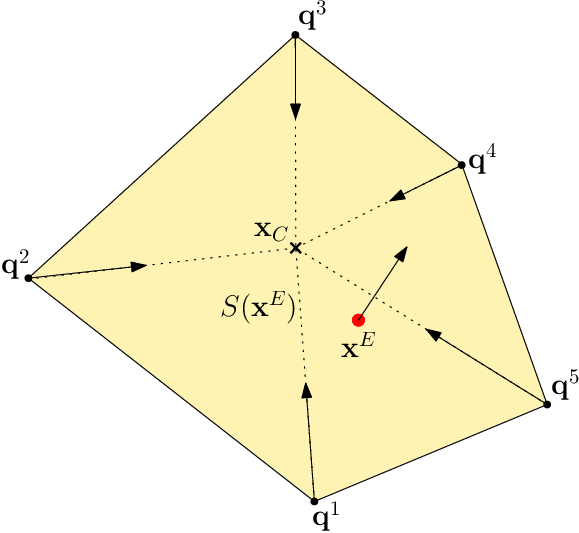}
    \caption{Motion policy for the vertices of $S(\mathbf{x}^E)$.}
    \label{fig:geom}
\end{figure}

Based on a geometric motion policy for the vertices of $S(\mathbf{x}^E)$, this section presents a control law for the pursuers. Consider $\mathbf{x}_{C} =[{x}_{\text{C}},{y}_{\text{C}}] $ denotes the centroid of $S(\mathbf{x}^E)$. The desired motion policy directs $i$th vertex of $S(\mathbf{x}^E)$ towards $\mathbf{x}_C$ and is expressed as follows:
\begin{equation}\label{eq:cl2}
    \begin{aligned} 
   \Dot{\mathbf{q}}^i= \begin{bmatrix}
        \Dot{q}^{i}_x\\
        \Dot{q}^{i}_y
    \end{bmatrix}= K \begin{bmatrix}
     {x}_{\text{C}}-q^{i}_x \\
     {y}_{\text{C}}-q^{i}_y     
    \end{bmatrix}, \; \forall \; i = 1,2,\dotsc, l,
    \end{aligned}
\end{equation}
where $K$ is a positive scalar. A geometric representation of the vertex motion policy is shown in Fig. \ref{fig:geom}.

\begin{theorem}\label{prop:inside_pt}
    For the motion policy proposed in Eq. (\ref{eq:cl2}), the evader area, $A_e$ decreases monotonically with time.
\end{theorem}

\begin{proof}
 Substituting for $\Dot{\mathbf{q}}_i$ from Eq. (\ref{eq:cl2}) in Eq. (\ref{eq:area_evader_dot}), the evader area dynamics can be expressed as

    \begin{align}
    \Dot{A}_e &= \frac{1}{2}\sum_{i =1}^l \left ( -K\frac{\partial A_e }{\partial q^i_x } ( q^{i}_x - {x}_{\text{C}}) - K\frac{\partial A_e }{\partial q^i_y } ( q^{i}_y - {y}_{\text{C}})\right ) \\
    &= \frac{-K}{2}\sum_{i =1}^l \bigl( (q^{i}_x-{x}_{\text{C}})( q^{i+1}_y - q^{i-1}_y) + (q^{i}_y -  {y}_{\text{C}})(q^{i-1}_x - q^{i+1}_x)\bigr)\\
     &= \frac{-K}{2}\sum_{i =1}^l \bigl( q^{i}_x( q^{i+1}_y - q^{i-1}_y) + q^{i}_y (q^{i-1}_x - q^{i+1}_x)\bigr)
    +\frac{K}{2}\sum_{i =1}^l \bigl({x}_{\text{C}}( q^{i+1}_y - q^{i-1}_y) +  {y}_{\text{C}}(q^{i-1}_x - q^{i+1}_x)\bigr)\label{eq:cl2_pr1}.  
    \end{align}
    Since $\sum_{i=1}^l  {x}_{\text{C}}( q^{i+1}_y - q^{i-1}_y)=\sum_{i=1}^l  {y}_{\text{C}}( q^{i-1}_x - q^{i+1}_x) =0$, Eq. (\ref{eq:cl2_pr1}) reduces to

    \begin{equation}\label{eq:cl2proof1}
        \begin{aligned}
            \Dot{A}_e &= \frac{-K}{2}\sum_{i =1}^l \left( q^{i}_x( q^{i+1}_y - q^{i-1}_y)+ q^{i}_y(q^{i-1}_x - q^{i+1}_x)\right).
        \end{aligned}
    \end{equation}
    Rearranging Eq. (\ref{eq:area_evader1}), the evader area can be expressed as
    \begin{align}
            A_e &= \frac{1}{4}  \bigl[2(q^1_x q^2_y + q^2_x q^3_y + \dotsc + q^l_x q^1_y  )  - 2(q^1_y q^2_x + q^2_y q^3_x + \dotsc + q^l_y q^1_x  )    \bigr]\\
&= \frac{1}{4}  \bigl[ q^1_x(q^2_y - q^l_y) + q^2_x(q^3_y - q^1_y) +\dotsc + q^l_x(q^1_y - q^{l-1}_y) +  q^1_y(q^l_x - q^2_x) + q^2_y(q^1_x - q^3_y) +\dotsc + q^l_y(q^{l-1}_x - q^1_x)\bigr]\\
&=\frac{1}{4} \sum_{i =1}^l \left(  q^{i}_x( q^{i+1}_y - q^{i-1}_y)+ q^{i}_y(q^{i-1}_x - q^{i+1}_x) \right)\label{eq:area_ev_manipulated}.
    \end{align}
    Using Eqs. (\ref{eq:cl2proof1}) and (\ref{eq:area_ev_manipulated}), the evader area dynamics is governed by
    \begin{equation}\label{eq:converge}
    \begin{aligned}
        \Dot{A}_e &= -2KA_e \\
        \implies A_e(t) &=A_e(0) \exp{(-2Kt)},         
    \end{aligned}
     \end{equation}
    where $A_e(0)$ is the area of $S(\mathbf{x}^E)$ at $t=0$.
\end{proof}

\begin{remark}
    Using Eq. (\ref{eq:converge}), it can be noted that the evader area decreases exponentially with time irrespective of the evader's evasion policy and speed.
\end{remark}

The pursuers' velocity input is obtained by substituting for $\Dot{\mathbf{q}}_i$ from Eq. (\ref{eq:cl2}) into Eq. (\ref{eq:allpursuer}) as  

\begin{equation}\label{eq:cl2_final}
   u^P_i  = \begin{cases}
$RHS of Eq. (\ref{eq:allpursuer}) where $ \Dot{\mathbf{q}_i} $ satisfies Eq. (\ref{eq:cl2})$ , &\text{if } \mathbf{{x}}^P_i \in \mathcal{M}\\
[0,0]^T, &\text{otherwise }
    \end{cases}
\end{equation}

\begin{proposition}
    For the proposed control law in Eq. (\ref{eq:cl2_final}), the capture of the evader is guaranteed.
\end{proposition}
\begin{proof}
    Let $R_{\min}$ denote the separation between the evader and its closest pursuer. By virtue of Voronoi partitioning, the pursuer's position is the mirror image of the evader's position about the edge shared between them. Accordingly,
    \begin{align}\label{eq:rmin}
        \begin{aligned}
            R_{\min} &= 2 \min \mid \mid \mathbf{x}^E - \mathcal{V}(\mathbf{x}^E) \mid \mid ,
        \end{aligned}
    \end{align}
     where $\mathcal{V}(x^E) = \{ \mathbf{x} : ||\mathbf{x} - \mathbf{x}^E|| = || \mathbf{x} - \mathbf{x}^P_i ||, ~ \forall \mathbf{x}^P_i \in \mathcal{M}   \}$. Using Eq. (\ref{eq:converge}), since $A_e \rightarrow 0$, the distance of the evader from the boundary of $S(\mathbf{x}^E)$, that is, $\min \mid \mid \mathbf{x}^E - \mathcal{V}(\mathbf{x}^E) \mid \mid \rightarrow 0$ and hence $R_{\min} \rightarrow 0 $. 
\end{proof}

While Proposition \ref{prop:inside_pt} presents a guarantee for target capture as $t \rightarrow \infty$, an upper bound on the capture time is deduced subsequently.

\begin{definition}
    For a convex polygon $\mathcal{P}$, the Chebyshev center is the center of the optimal circle or the largest circle drawn within $\mathcal{P}$ and the Chebyshev radius is the radius of that circle.  
\end{definition}
\begin{property}\label{prop:cheb_center}
    \cite{boyd2004convex} The Chebyshev center $\mathbf{x}_{cb}$ is also the point farthest from the boundary of $\mathcal{P}$.
\end{property}
\begin{property}\label{prop:cir_poly}
     \cite{polya1955more} In the family of all $m-$sided convex polygons containing a given circle, the regular polygon that circumscribes the circle has the least area.
\end{property}

\begin{lemma}\label{lemma:1}
    Consider two $m-$sided convex polygons ($m>2$), $\mathcal{P}_1$ and $\mathcal{P}_2$ with equal areas where $\mathcal{P}_1$ is regular and $\mathcal{P}_2$ is irregular. The Chebyshev radius of $\mathcal{P}_1$ is greater than that of $\mathcal{P}_2$.
\end{lemma}

\begin{proof}
    Let the Chebyshev radius of $\mathcal{P}_1$ and $\mathcal{P}_2$ be $r_1$ and $r_2$, respectively. Using Property \ref{prop:cir_poly}, there exists an $m-$sided regular polygon $\mathcal{P}_3$ which circumscribes a circle of radius $r_2$ but has a lesser area than $\mathcal{P}_2$ as illustrated in Fig. \ref{fig:cheb_radi}. Consider $A_{\mathcal{P}_1}$, $A_{\mathcal{P}_2}$ and $A_{\mathcal{P}_3}$ are the areas of the polygons $\mathcal{P}_1$, $\mathcal{P}_2$ and $\mathcal{P}_3$, respectively. Using the relation between the area of a regular polygon and the radius of its inscribed circle \cite{mathopenref_polygon_area}, the radii $r_1$ and $r_2$ are given by
    \begin{align}\label{eq:rin_ch}
        r_{1} =\sqrt{\dfrac{A_{\mathcal{P}_1}}{m \tan\left( \dfrac{\pi}{m} \right)} },~ ~~~& r_{2} =\sqrt{\dfrac{A_{\mathcal{P}_3}}{m \tan\left( \dfrac{\pi}{m} \right)} }.
    \end{align}
    Since $A_{\mathcal{P}_1}=A_{\mathcal{P}_2}>A_{\mathcal{P}_3}$, using Eq. (\ref{eq:rin_ch}), the Chebyshev radius of $\mathcal{P}_1$ is greater than that of $\mathcal{P}_2$, that is, $r_1>r_2.$

    \begin{figure*}[!hbt]
    \centering
    \includegraphics[width=0.7\textwidth]{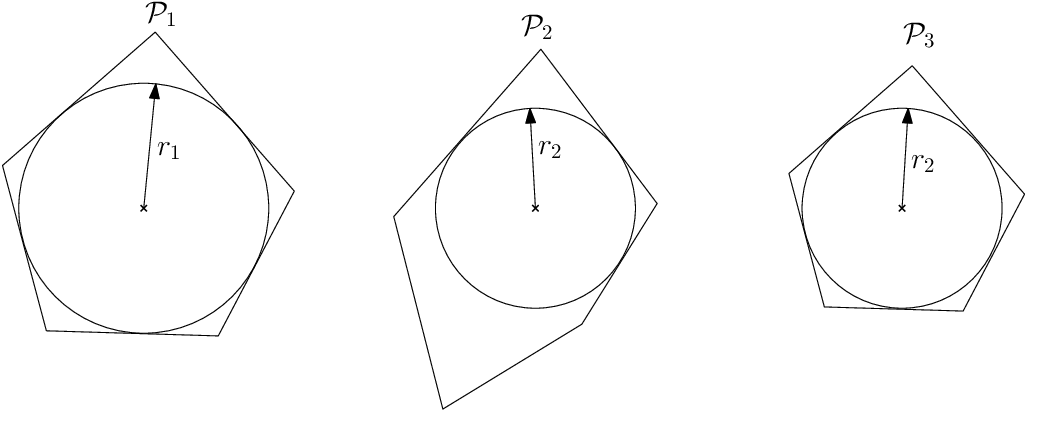}
    \caption{Comparison of the Chebyshev radii of $\mathcal{P}_1,\mathcal{P}_2,\mathcal{P}_3$ ($A_{\mathcal{P}_1}=A_{\mathcal{P}_2}>A_{\mathcal{P}_3}$).}
    \label{fig:cheb_radi}
\end{figure*}
\end{proof}

\begin{figure}[!hbt]
    \centering
    \includegraphics[width=0.45\textwidth]{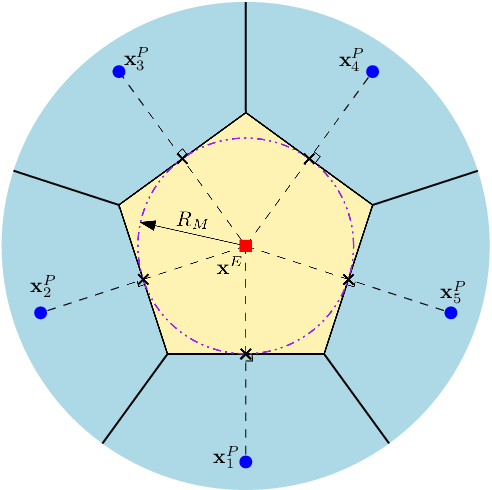}
    \caption{Largest possible circle drawn in $S(\mathbf{x}_E)$ considering $\mathbf{x}_E$ as the Chebyshev center.}
    \label{fig:cheb}
\end{figure}

Using Lemma \ref{lemma:1}, for a specific value of $A_e$, the Chebyshev radius is maximum when $S(\mathbf{x}^E)$ is a regular polygon. Further, the distance between the evader and a pursuer is twice that between the evader and the corresponding shared side of $S(\mathbf{x}_E)$. To find the upper bound on the capture time, subsequent analysis considers the following assumptions:
\begin{enumerate}[label=(A\arabic*),resume, start=5]
     \item The evader always lies at the Chebyshev center of $S(\mathbf{x}_E)$.\label{assum:5}
       \item $S(\mathbf{x}_E)$ is considered as an $m-$sided regular polygon at all times. \label{assum:6}
\end{enumerate}
Fig. \ref{fig:cheb} depicts the largest possible circle drawn in $S(\mathbf{x}^E)$ of area $A_e$ with $R_M$ being the Chebyshev radius of $S(\mathbf{x}^E)$. Using Eq. (\ref{eq:rmin}) and Property \ref{prop:cheb_center},
\begin{equation}\label{eq:rmin_rm}
    R_{\min} = 2R_M.
\end{equation}

\begin{proposition}\label{prop:time_int}
    For the proposed policy, the upper bound on the time of evader capture is 
    \begin{equation}\label{eq:tfrc}
    t_C^U = -\dfrac{1}{2K}\log \left (\dfrac{mr_c^2}{4A_e(0)} \tan \left( \dfrac{\pi}{m}\right )\right ).
\end{equation}
\end{proposition}
\begin{proof}
   Using the assumptions \ref{assum:5}, \ref{assum:6} and Eq. (\ref{eq:rin_ch}), the Chebyshev radius is expressed in terms of $A_e(t)$ as follows:
\begin{align}\label{eq:rm1}
    R_M(t) &= \sqrt{\dfrac{A_e(t)}{m \tan\left( \dfrac{\pi}{m} \right)} }.
\end{align}
Using Eqs. (\ref{eq:converge}) and (\ref{eq:rm1}),
\begin{equation}\label{eq:rm2}
    R_M(t) = \sqrt{\dfrac{A_e(0) \exp{(-2Kt)}}{m \tan\left( \dfrac{\pi}{m} \right)} }.
\end{equation}
Using Eqs. (\ref{eq:capture}) and (\ref{eq:rmin_rm}), the capture criteria can be expressed in terms of $R_M$ as
\begin{equation}\label{eq:final_capture}
   r_c =2 R_M.
\end{equation}
Using Eqs. (\ref{eq:rm2}) and (\ref{eq:final_capture}),
\begin{align}
   \sqrt{\dfrac{A_e(0) \exp{(-2Kt_C^U)}}{m \tan\left( \dfrac{\pi}{m} \right)} } &= \dfrac{r_c}{2},\\
   \implies  t_C^U &= -\dfrac{1}{2K}\log \left (\dfrac{mr_c^2}{4A_e(0)} \tan \left( \dfrac{\pi}{m}\right )\right ).
\end{align}
\end{proof}


\section{Simulation Results}\label{sec:5}

To demonstrate the characteristics of the proposed motion policy, this section presents simulation results considering 3 pursuers. The capture radius $r_C =0.2$ m. The constant gain used in (\ref{eq:cl2}) satisfies $K =0.05$. In all simulations, the pursuers' initial positions are (-80.65 m, 44.48 m), (63.63 m, -70.02 m) and (63.51 m, 31.92 m), and the evader's initial position is (0.7438 m, 18.92 m). For the resulting Voronoi partition at $t=0$, $A_e(0)=13370$ m$^2$ and $m =3$. Accordingly,  the upper bound on the time of evader capture calculated using Eq. (\ref{eq:tfrc}) is 124.58 s.

\subsection{ Case 1: Non-maneuvering Evader}\label{subsec:mn}

\begin{figure}[!hbt]
    \centering
    \includegraphics[width = 0.5\textwidth]{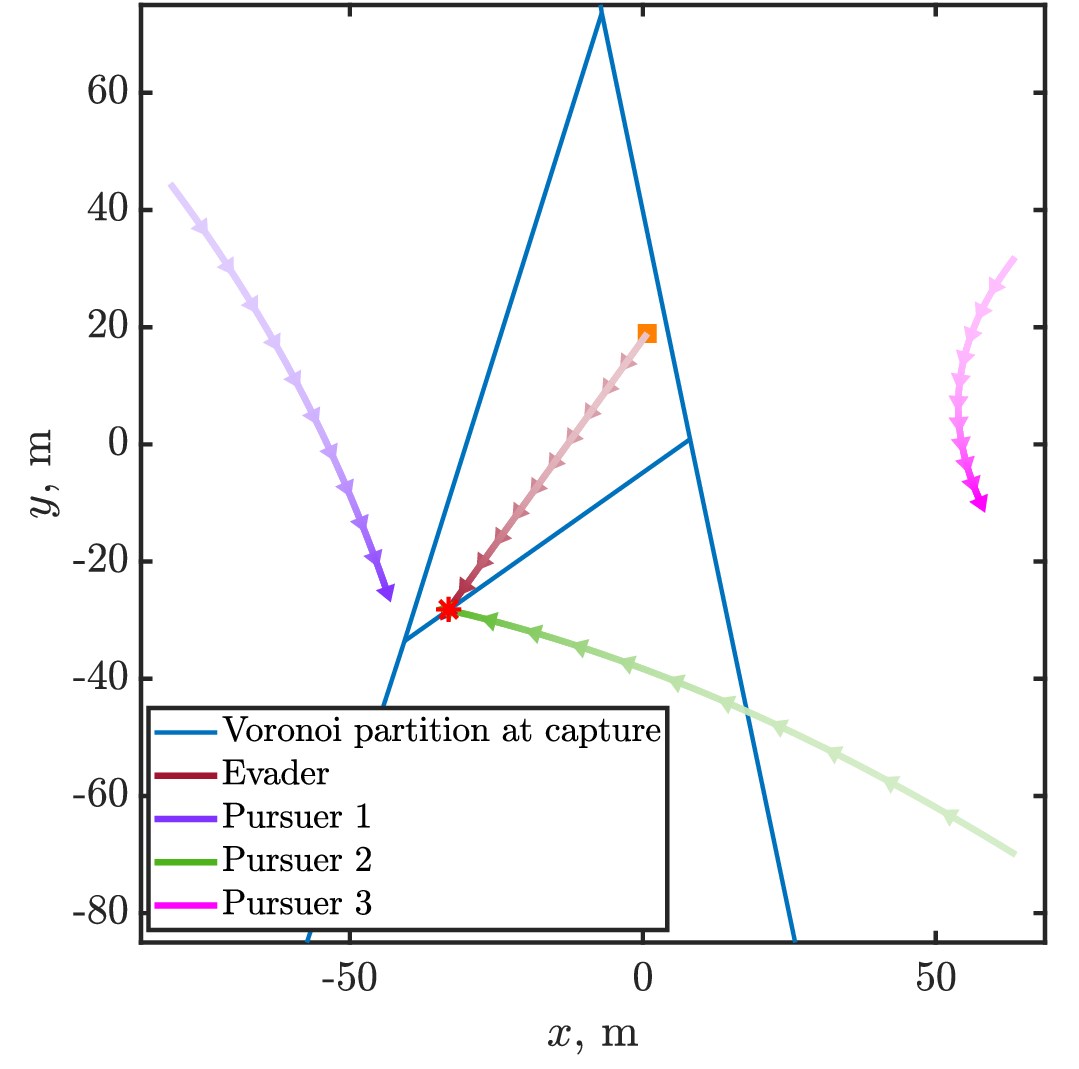}
    \caption{Case 1: Agents' trajectory .}
    \label{fig:st_traj}
\end{figure}

\begin{figure*}[!hbt]
    \centering
    \subfloat[\centering Evader area.]{{\includegraphics[width=0.45\textwidth]{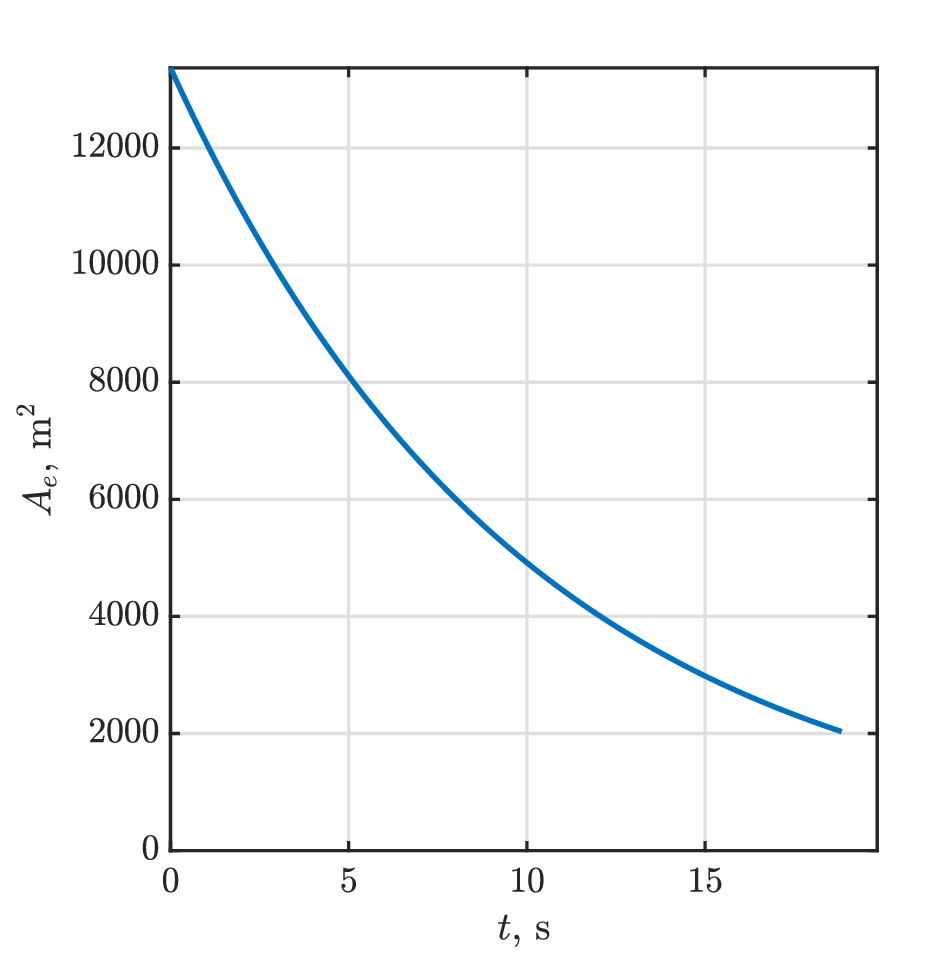} }}
    \quad
    \subfloat[\centering Evader separation distance from its closest pursuer.]{{\includegraphics[width=0.45\textwidth]{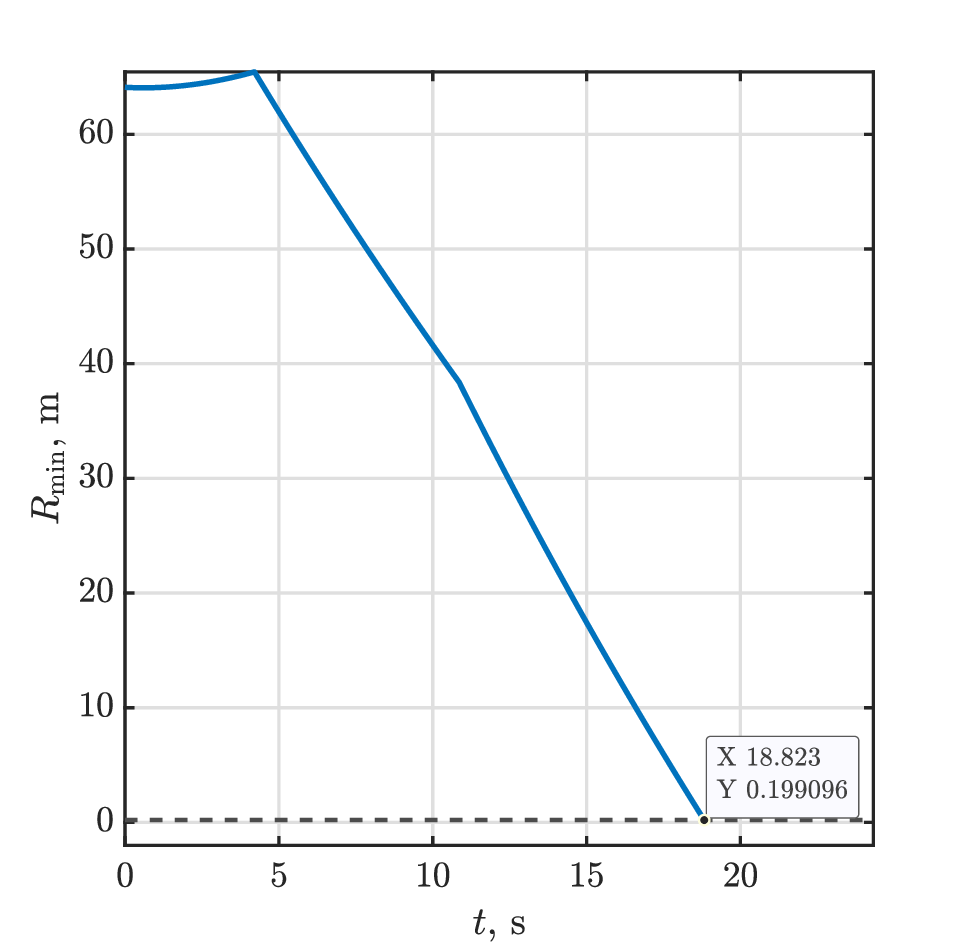} }}
    \caption{Case 1: Evader capture parameters.}
    \label{fig:st_traj_char}
\end{figure*}

\begin{figure}[!hbt]
    \centering
    \includegraphics[width = 0.5\textwidth]{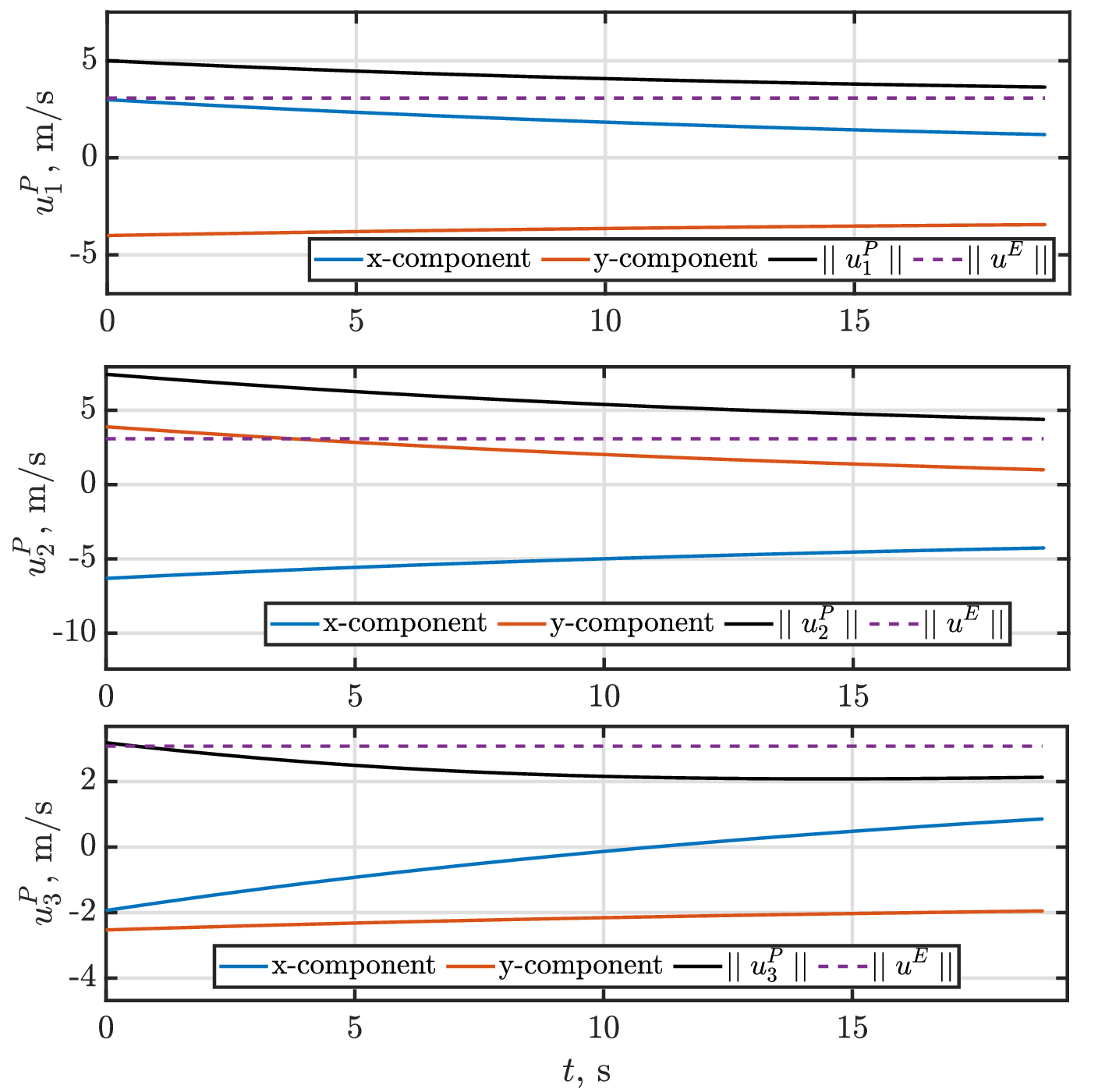}
    \caption{Case 1: Pursuers' velocity profile. }
    \label{fig:st_traj_velprof}
\end{figure}

Here, the evader follows a straight line path with constant velocity, $u^E  = [-1.8,-2.5]$ m/s. The trajectory plot for the agents is shown in Fig. \ref{fig:st_traj}. The colour of solid lines depicting agents' trajectory darkens with time which shows the spatio-temporal information of their position. Square and asterisk markers depict the start and end position of the evader, respectively. As shown in Fig. \ref{fig:st_traj_char}a, the evader area exponentially decreases with time. The separation between the evader and the closest pursuer in Fig. \ref{fig:st_traj_char}b shows that the evader is captured at $t = 18.823~$s. The speed profile of the pursuers is shown in Fig. \ref{fig:st_traj_velprof}. In this case, it can be observed that Pursuer 1 and Pursuer 2 move at speeds faster than that of the evader. However, the speed of Pursuer 3 remains lower than that of the evader for $t \in [0.6,18.823]$ s.

\subsection{Case 2: Maneuvering Evader}\label{subsec:m}

\begin{figure}[!hbt]
    \centering
    \includegraphics[width = 0.5\textwidth]{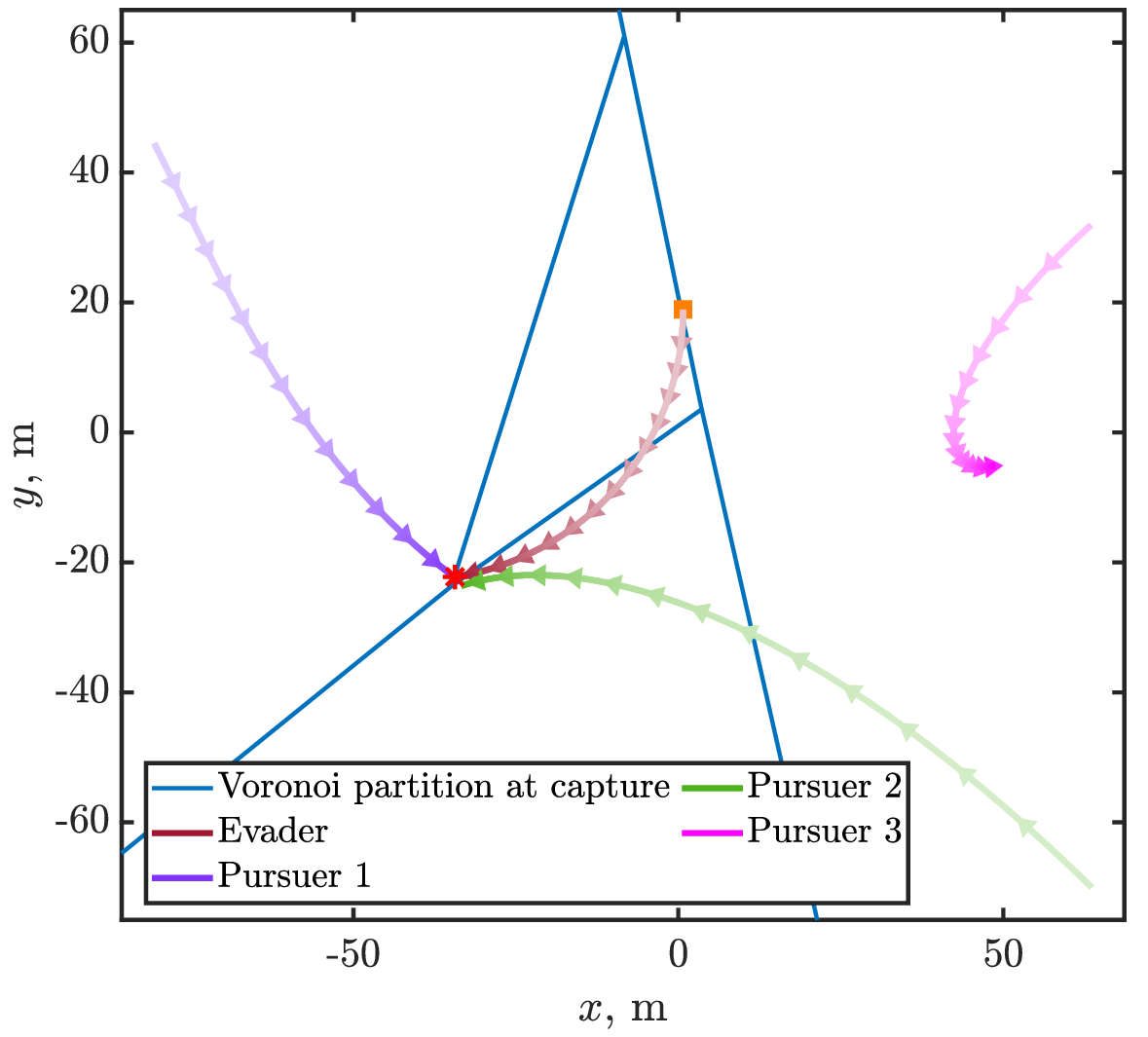}
    \caption{Case 2: Agents' trajectory.}
    \label{fig:cir_traj}
\end{figure}

\begin{figure*}[!hbt]
    \centering
    \subfloat[\centering Evader area.]{{\includegraphics[width=0.45\textwidth]{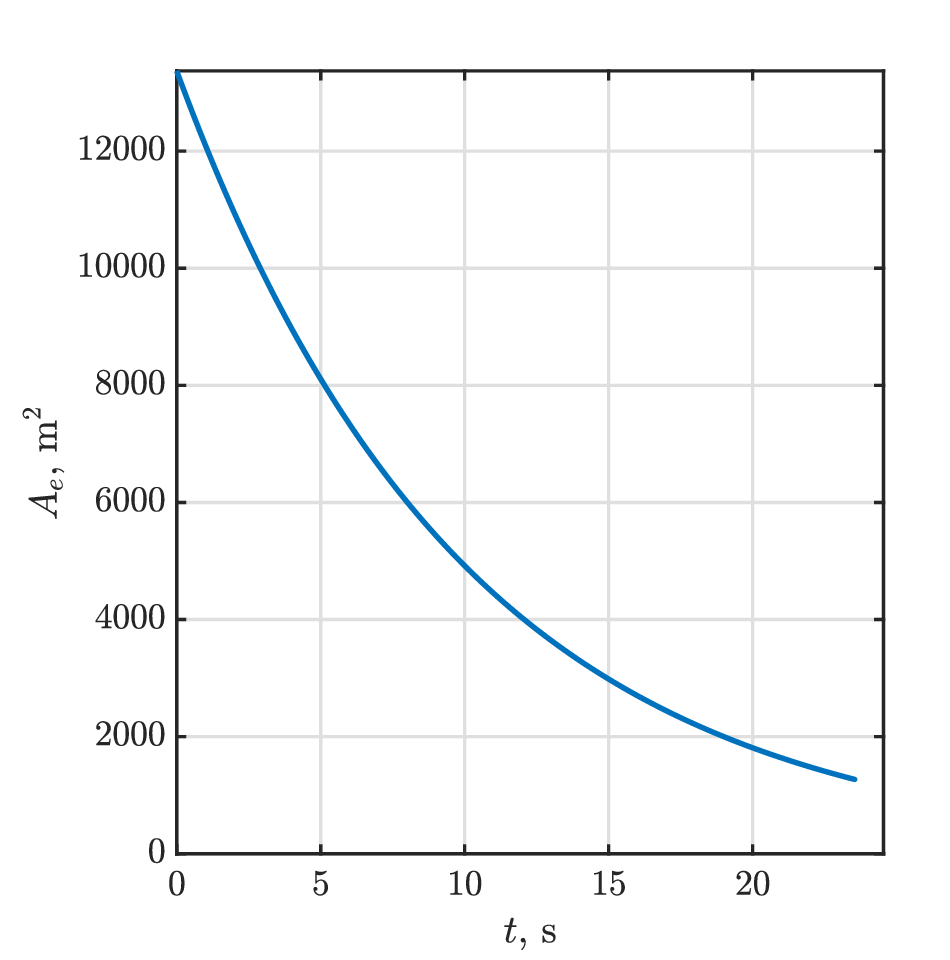} }}%
    \quad
    \subfloat[\centering Evader separation distance from its closest pursuer.]{{\includegraphics[width=0.45\textwidth]{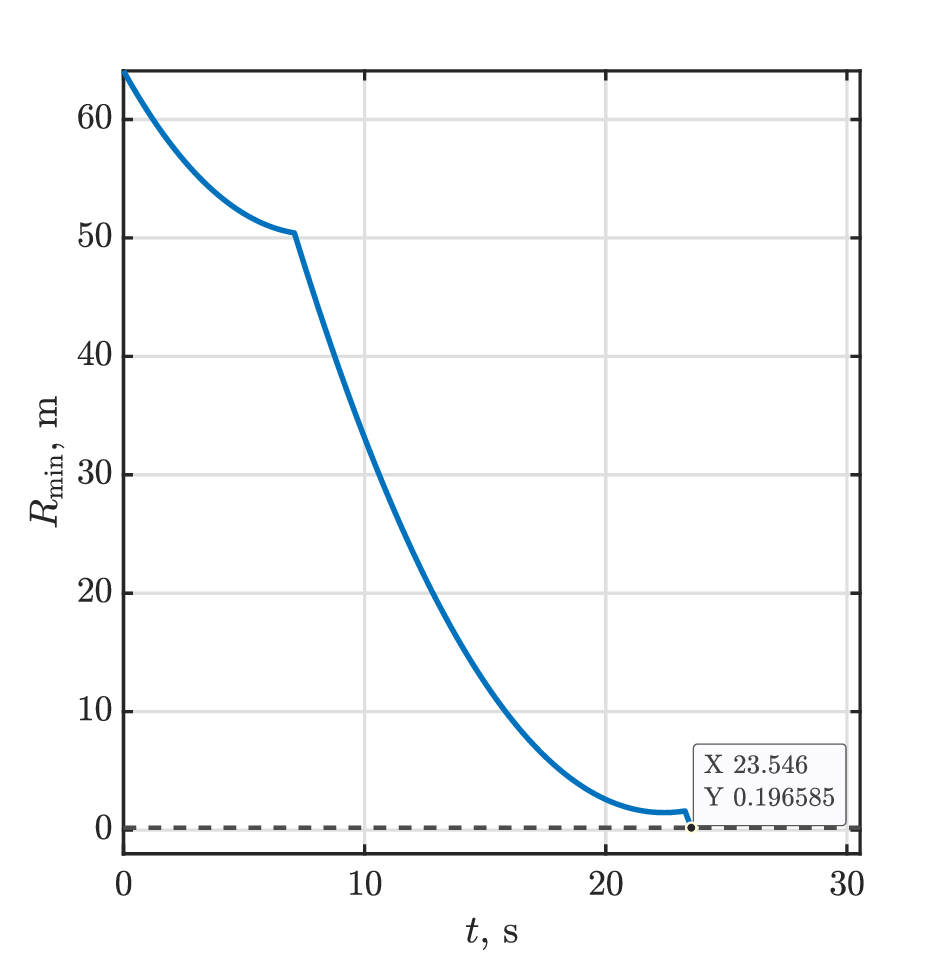} }}%
    \caption{Case 2: Evader capture parameters.}%
    \label{fig:mnvr_traj_char}%
\end{figure*}

In this case, the velocity profile of the evader is considered as $u^E = [-2.5\sin(0.06t),-2.5\cos(0.06t)]^T$ m/s. The trajectory plot for the agents in Fig. \ref{fig:cir_traj} shows that the evader is captured by Pursuer 1. The area of the evader's proximity region is shown to shrink exponentially with time in Fig. \ref{fig:mnvr_traj_char}a. The distance between the evader and its closest pursuer in Fig. \ref{fig:mnvr_traj_char}b shows that the capture happens at $t = 23.546~$s. Further, two specific evasion policies are considered, and the capture time is analyzed for each of these policies.
\begin{enumerate}[label=Policy (\arabic*)]
    \item Evader moves to the centroid of $S(\mathbf{x}^E)$ \cite{pierson2016intercepting}: The evader's control input is given as $u^E = 0.2(\mathbf{x}_C -\mathbf{x}^E)$. The resulting capture time of the evader $t_C=$ 114.67 s. \label{policy1}
    \item Evader moves to the Chebyshev center of $S(\mathbf{x}^E)$: Here, $u^E = 0.2(\mathbf{x}_{cb} -\mathbf{x}^E)$ and $t_C=$ 118.3 s. \label{policy2}
\end{enumerate}
It can be seen that the use of \ref{policy1} or \ref{policy2} increases the capture time. However, the capture time is lower than the upper bound $t_C^U=124.58$ s as obtained through the analytic result in Eq. (\ref{eq:tfrc}).

\begin{figure}[!hbt]
    \centering
    \includegraphics[width = 0.5\textwidth]{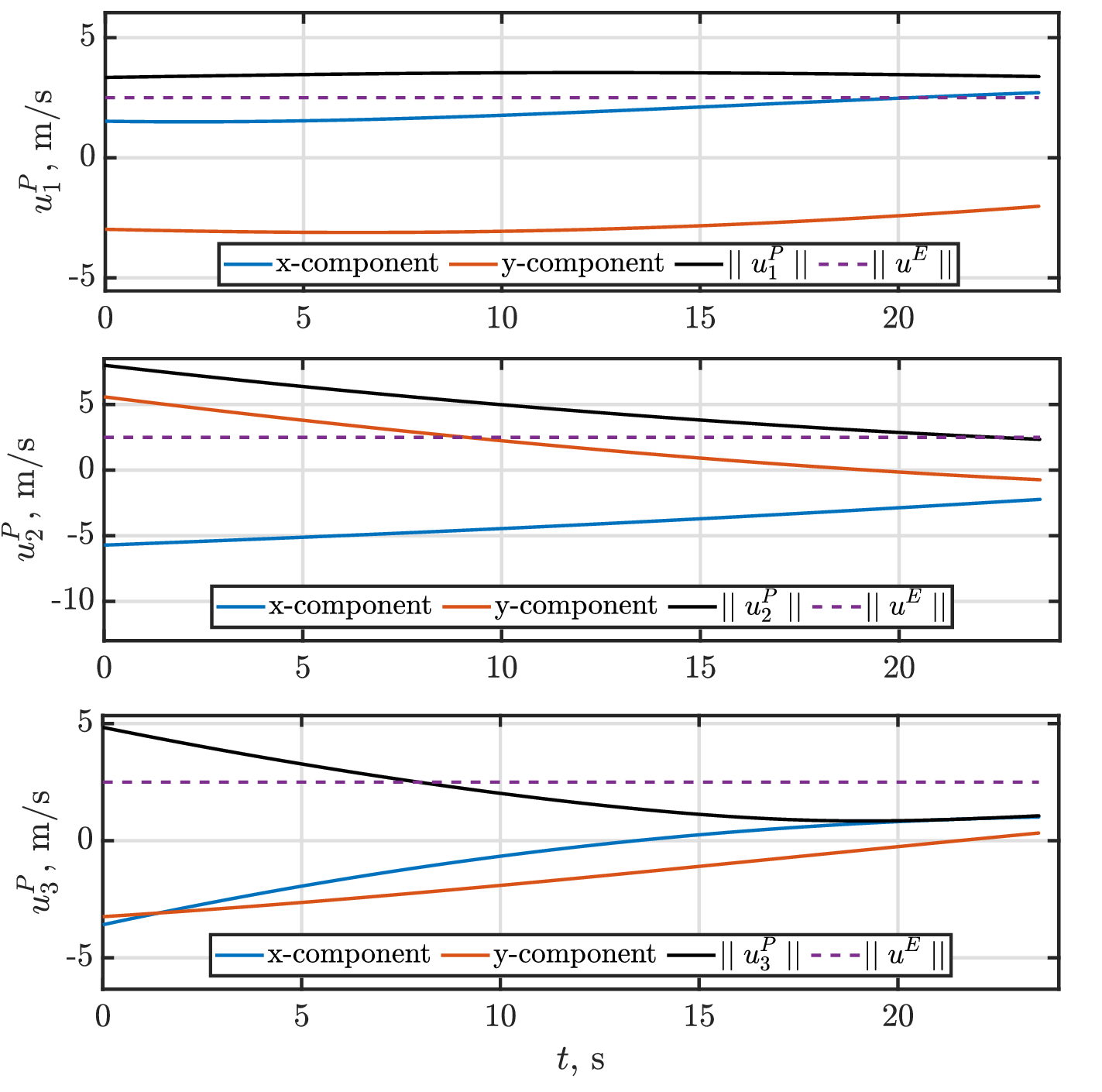}
    \caption{Case 2: Pursuers' velocity profile. }
    \label{fig:cir_traj_velprof}
\end{figure}

\subsection{Case 3: Maneuvering Target Capture using Noisy Position Information}

As per Eq. (\ref{eq:allpursuer}), the pursuers' velocity input is computed while assuming that the evader's position and velocity information is known. In this case study, we consider a realistic scenario where the evader's position is the only information available. Kalman filtering techniques are widely used in the estimation of unknown system states of a target from noisy measurement data \cite{zarchan2010boost,lawton1998comparison,zarchan2005progress}. To demonstrate the effectiveness of the proposed strategy in such a scenario, a discrete second-order polynomial Kalman filter \cite{zarchan2005progress} is used hereinafter to estimate the evader's position and velocity information from its noisy position information.

\subsubsection{Second-order Polynomial Kalman Filter}

 Consider the evader state $\textit{\textbf{x}}^E_k = [x^E_k,\dot{x}^E_k,\ddot{x}^E_k,y^E_k,\dot{y}^E_k,\ddot{y}^E_k]$ at time step $k$. In the case of no deterministic disturbance and control input, the discrete Kalman-filtering equation is given by
\begin{align}\label{eq:filt_evad}
    \hat{\textit{\textbf{x}}}^E_{k}  = \Phi_k \hat{\textit{\textbf{x}}}^E_{k-1} + {\textit{\textbf{K}}}_k ({\textit{\textbf{z}}}_k- {\textit{\textbf{H}}}\Phi_k \hat{\textit{\textbf{x}}}^E_{k-1}),
\end{align}
where $\hat{\textit{\textbf{x}}}^E_{k}$ is the estimated evader state, $\Phi_k$ is the state transition matrix, ${\textit{\textbf{K}}}_k$ is the Kalman gain matrix, ${\textit{\textbf{z}}}_k$ is the measurement vector containing noisy position information and ${\textit{\textbf{H}}}$ is the measurement matrix. Here,
\begin{align}
     \Phi_k =\begin{bmatrix}
        1 & \Delta t &0.5\Delta t^2&0&0&0\\
        0 & 1 & \Delta t & 0 & 0 & 0 \\
        0 & 0 & 1 & 0 & 0 & 0 \\
         0&0&0 &1 & \Delta t &0.5\Delta t^2\\
        0&0&0 & 0 & 1 & \Delta t \\
        0 & 0 & 0 & 0 & 0 & 1 
    \end{bmatrix},~ {\textit{\textbf{H}}} = \begin{bmatrix}
         1& 0 & 0 &0 &0 &0  \\
         0& 0 & 0 &1 &0 &0
    \end{bmatrix},
\end{align}
where $\Delta t$ is the time step size. The Kalman gain matrix is computed using a set of recursive matrix equations called Riccati equations given by
\begin{align}
   {\textit{\textbf{M}}}_k &= \Phi_k {\textit{\textbf{P}}}_{k-1}\Phi_k^T +{\textit{\textbf{Q}}}_k \\
   {\textit{\textbf{K}}}_k &= {\textit{\textbf{M}}}_k{\textit{\textbf{H}}}^T \left ( {\textit{\textbf{H}}} {\textit{\textbf{M}}}_k {\textit{\textbf{H}}}^T + {\textit{\textbf{R}}}_k \right )^{-1}\\
   {\textit{\textbf{P}}}_k &= \left ({\textit{\textbf{I}}} -{\textit{\textbf{K}}}_k {\textit{\textbf{H}}} \right ) {\textit{\textbf{M}}}_k,
\end{align}
where ${\textit{\textbf{P}}}_k$ is the error covariance matrix after an update, ${\textit{\textbf{M}}}_k$ is the error covariance matrix before an update, ${\textit{\textbf{R}}}_k$ is the measurement noise matrix, and ${\textit{\textbf{Q}}}_k$ is the process noise matrix. The estimated evader position and velocity information, that is, $[\hat{x}^E_k,\hat{\dot{x}}^E_k,\hat{y}^E_k,\hat{\dot{y}}^E_k]$ as obtained from Eq. (\ref{eq:filt_evad}) is used in Eq. (\ref{eq:allpursuer}) to determine pursuers' velocity inputs.

\subsubsection{Simulation Scenario}
\begin{figure}[!hbt]
    \centering
    \includegraphics[width = 0.5\textwidth]{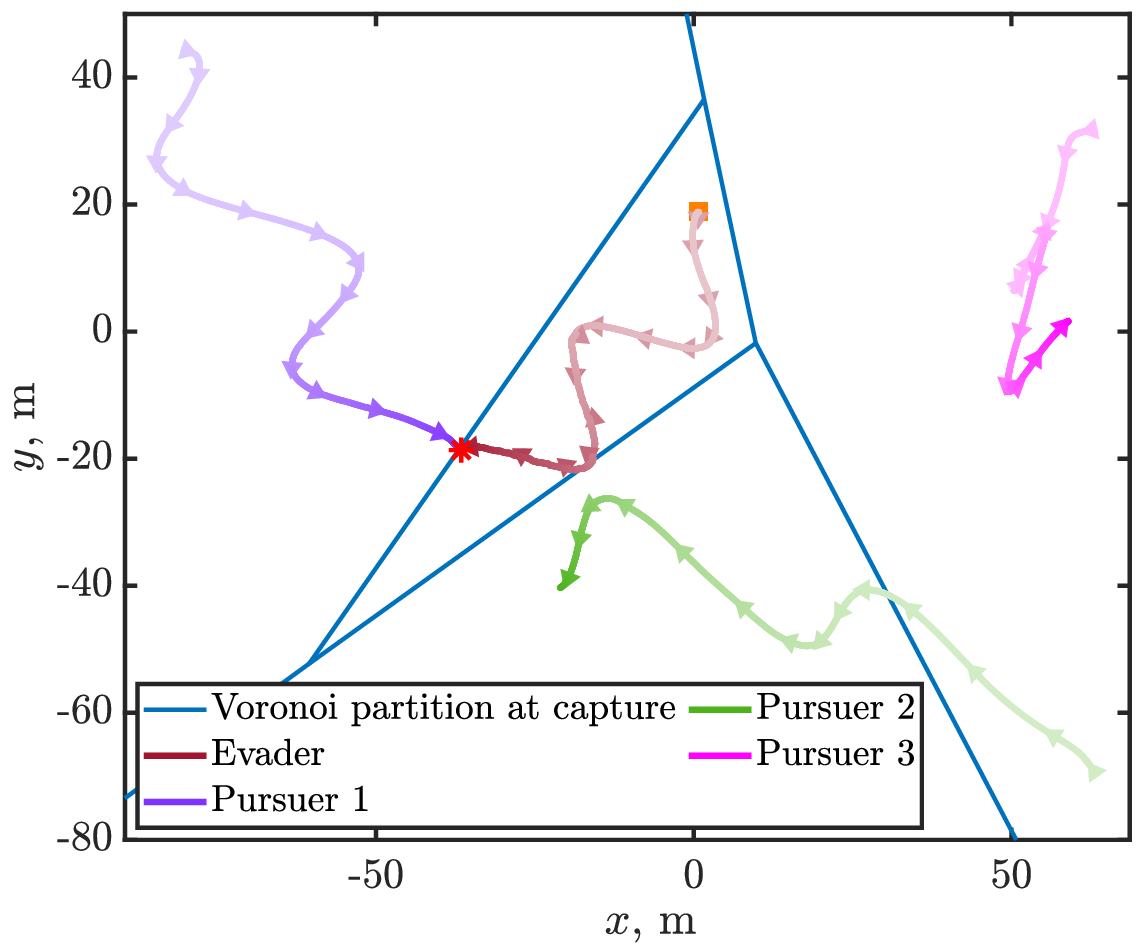}
    \caption{Case 3: Agents' trajectory.}
    \label{fig:noise_traj}
\end{figure}

\begin{figure*}[!hbt]
    \centering
    \subfloat[\centering Evader area.]{{\includegraphics[width=0.45\textwidth]{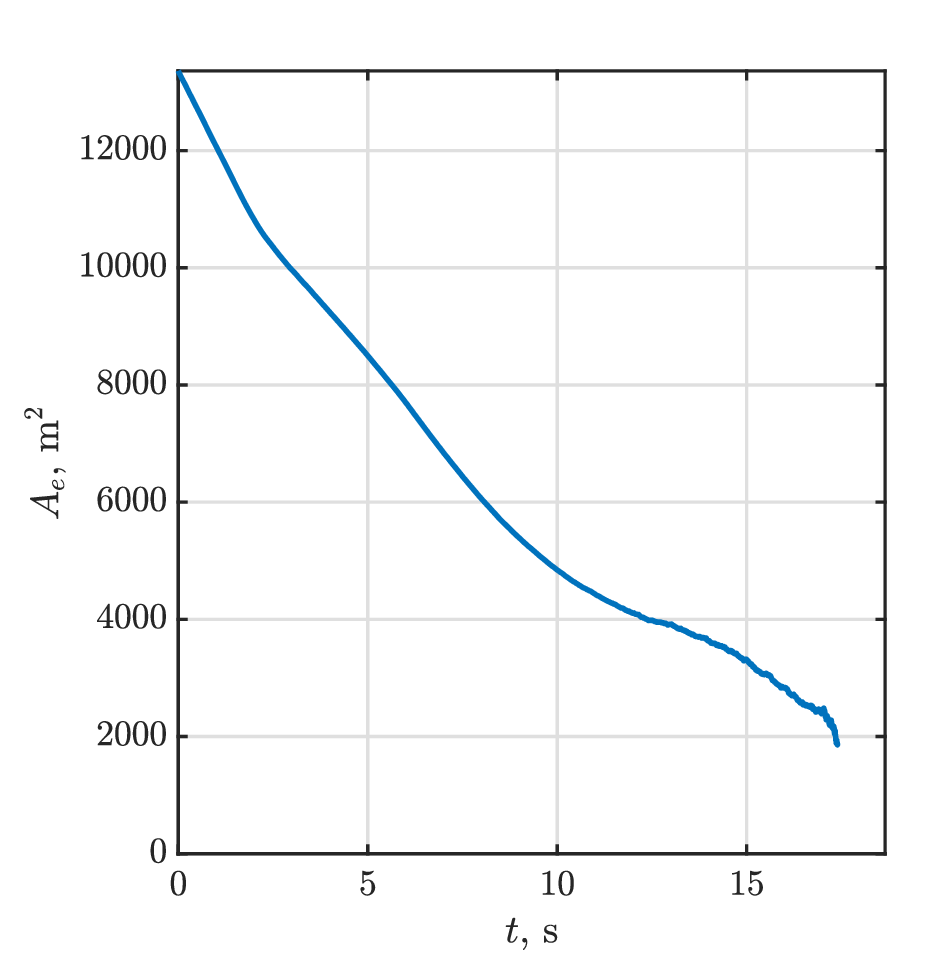} }}%
    \quad
    \subfloat[\centering Evader separation distance from its closest pursuer.]{{\includegraphics[width=0.45\textwidth]{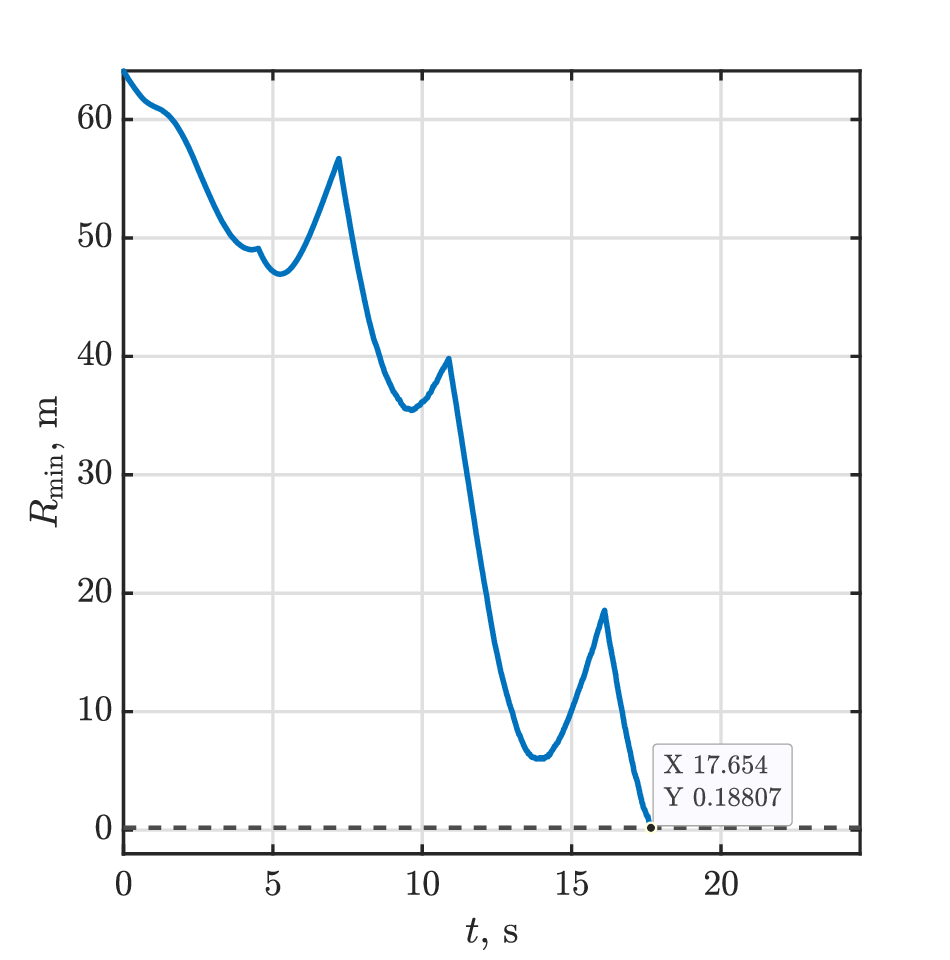} }}%
    \caption{Case 3: Evader capture parameters.}%
    \label{fig:noise_traj_char}%
\end{figure*}

\begin{figure}[!hbt]
    \centering
    \includegraphics[width = 0.5\textwidth]{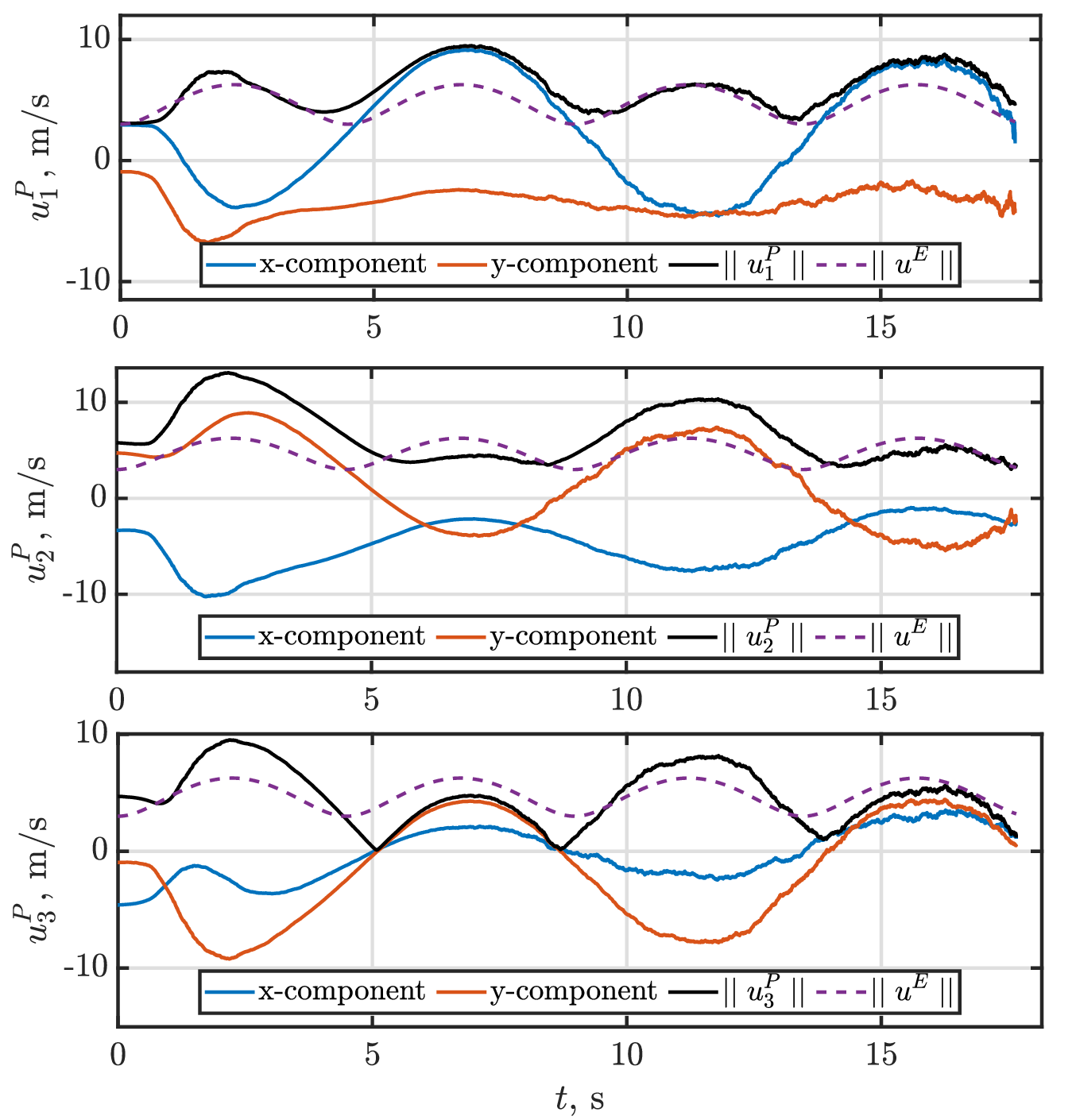}
    \caption{Case 3: Pursuers' velocity profile. }
    \label{fig:noise_traj_velprof}
\end{figure}

In this example, the evader follows a sinusoidal path governed by $u^E = [3.89\sin(0.7t)-2.12,-3.89\sin(0.7t)-2.12]^T$ m/s. To incorporate white Gaussian noise in the position information of the evader, \textit{awgn} function is utilized with the signal-to-noise ratio of 20 dB in \textsc{Matlab}\textsuperscript{\circledR}. The trajectory plot in Fig. \ref{fig:noise_traj} shows that the evader is captured by Pursuer 1. The variation of the area of the evader's proximity region with time is shown in Fig. \ref{fig:noise_traj_char}a. The time evolution of the separation between the evader and its closest pursuer in Fig. \ref{fig:noise_traj_char}b shows that the time of evader capture is 17.654 s. The velocity profile for all pursuers is shown in Fig. \ref{fig:noise_traj_velprof}. 

\section{Conclusion}\label{sec:6}

This paper addresses the multiple pursuers-single evader problem in an unbounded region where it is assumed that the evader's initial position lies inside the convex hull formed by the pursuers' initial position. The evader's proximity region is defined using the Voronoi partition of the plane containing the agents. A direct relationship between the velocity and position of active pursuers, evader, and vertices of the evader's proximity region is derived. The proposed motion policy directs the vertices of the evader's proximity region towards its centroid, and the corresponding velocity inputs for the active pursuers is deduced. The proposed approach guarantees an exponential reduction in the area of the evader's proximity region regardless of its motion policy. An upper bound on the time of evader capture is deduced using the limiting geometry of the evader's proximity region and its Chebyshev radius. Simulation results illustrate that, following the proposed approach, it is not necessary for all pursuers to move faster than the evader. Further, a second-order polynomial Kalman filter is utilized in the scenario where only noisy position information of the evader is available. Future research direction can be towards exploring this strategy for capturing multiple evaders in a three-dimensional environment.

\bibliography{updated_sample}

\end{document}